\documentstyle{article}
\begin{document}
\title{Bound state equation for 4 or more relativistic particles.} 
\author{ J. Bijtebier\thanks{Senior Research Associate at the
 Fund for Scientific Research (Belgium).}\\
 Theoretische Natuurkunde, Vrije Universiteit Brussel,\\
 Pleinlaan 2, B1050 Brussel, Belgium.\\ Email: jbijtebi@vub.ac.be}
\maketitle
\begin{abstract}   \noindent
We apply the 3D reduction method we recently proposed for the N-particle
Bethe-Salpeter equation to the 4-particle case. We find that the writing of the
$\,N\!\ge\!4\,$ Bethe-Salpeter equation is not a straightforward task, owing to
the presence of mutually unconnected interactions, which could lead to an
overcounting of some diagrams in the resulting full propagator. We overcome
this difficulty in the $\,N=4\,$ case by including three counterterms in the
Bethe-Salpeter kernel. The application of our 3D reduction method to the
resulting Bethe-Salpeter equation suggests us a modified 3D reduction
method, which gives directly the 3D potential, without the need of writing the
Bethe-Salpeter kernel explicitly. The modified reduction method is usable for
all $\,N.$     
\end{abstract}
electrodynamics.
Breit's equation.\par  Relativistic bound states. Relativistic wave equations.
\\\\
 \newpage
\tableofcontents

\section{Introduction.} The Bethe-Salpeter equation \cite{1,2} is the usual tool
for computing relativistic bound states [1-20]. The principal difficulty
of this equation comes from the presence of N-1 (for N particles) unphysical
degrees of freedom: the relative time-energy degrees of freedom. In a recent
work \cite{20}, we built a 3D reduction of the two-fermion Bethe-Salpeter
equation around an unspecified instantaneous approximation of the Bethe-Salpeter
kernel, the difference with the exact kernel being corrected in a series of
higher-order contributions to the final 3D potential. This potential being not
explicitly hermitian, we performed a second symmetrizing series expansion at the
3D level. The result of the combination of both series gave us an hermitian
potential which turned out to be independent of the starting instantaneous
approximation, and was in fact a compact expression of the potential that
Phillips and Wallace
\cite{19} compute order by order. We found that this 3D potential was also
obtainable directly by starting with an approximation of the free propagator,
based on integrals in the relative energies instead of the more usual
$\,\delta-$constraint (integrating propagator-based reduction). In this form, the
method was easily generalizable to a system of N particles, consisting of any
combination of bosons and fermions. Special cases of N=2 or 3-fermion systems
were examined. Taking the retarded part of the full propagator at equal times,
following Logunov and Tavkhelidze
\cite{4} (for two particles) and Kvinikhidze and Stoyanov \cite{KS} (for
three particles) also leads to the same 3D equations.   
\par
For more detailed explanations about our integrating propagator-based reduction,
about related problems like Lorentz covariance, cluster separability or
continuum dissolution, or for comparisons with other approaches, we refer to
our preceding work \cite{20}.\par 

 In the present work we test our reduction method in the four-body problem. The
first step is of course the writing of the starting Bethe-Salpeter equation
itself. It is an homogeneous equation for a Bethe-Salpeter amplitude, which is
derived by identifying the residues of the bound state poles in the
inhomogeneous Bethe-Salpeter equation for the full propagator $\,G.\,$ The
interactions are introduced via the Bethe-Salpeter kernel $\,K,\,$ which is such
that the inhomogeneous Bethe-Salpeter equation reproduces, by iterations, the
full propagator $\,G\,$ as deduced by the Feynman's graphs method. The writing
of the Bethe-Salpeter kernel
$\,K\,$ is straightforward in the two and three-body problems. In the
four-body problem, however, we meet a new difficulty due to the mutual
unconnectedness of the three pairs of two-body irreducible kernels (12)(34),
(13)(24) and (14)(23): adding simply the six two-body
irreducible kernels (plus the four irreducible three-body
kernels and the irreducible four-body kernel) would lead to an
overcounting of some graphs in the expansion of $\,G.\,$ We would get for
example the sequences of kernels (12)(34) and (34)(12), which are in
fact a unique graph or part of graph (unlike sequences like (12)(23) and
(23)(12) which are different). It is however not very difficult to
correct this overcounting by including three counterterms in $\,K.\,$
Computing then the first terms of the 3D potential by our 3D reduction
method, we find that, at the often used approximation in which we keep
only the two-body kernels, the (relative and total) energy-independent part of
these kernels and the positive-energy part of
the propagators, the final 3D potential is simply the sum of the six
two-particle potentials.\par
It is clear that the difficulty of the overcounting of the sequences
of unconnected kernels will not be so easily overcomed for
$\,N\!\ge\!5.\,$ This suggested us to modify our 3D reduction
method in order to avoid the explicit writing of
the Bethe-Salpeter kernel $\,K.\,$ Our 3D potential is now written directly in a
straightforward way for any number of particles, with simply the additional
prescription of removing the duplicating diagrams which appear when
$\,N\!\ge\!4.\,$ 
\par
\hfill\break
In section 2, we recall our 3D reduction method of \cite{20}
for the N-body Bethe-Salpeter equation. In section 3 we apply it to
the four-body problem. In section 4 we suggest a general 3D
reduction method starting directly from the expansion of the full
propagator $\,G,\,$ as given by Feynman graphs, without the need of
writing explicitly the kernel $\,K\,$ of the Bethe-Salpeter
equation. Section 5 is devoted to conclusions.

\section{Integrating propagator-based reduction for the N-body Bethe-Salpeter
equation.}
 The inhomogeneous and homogeneous Bethe-Salpeter equations can be written
\begin{equation}G\,=\,G^0\,+\,G^0KG\label{1}\end{equation}
\begin{equation}\Phi\,=\,G^0K\,\Phi\label{2}\end{equation}
where $\,G\,$ is the full propagator and $\,\Phi\,$ the Bethe-Salpeter
amplitude. The free propagator $\,G^0\,$ for a system of
$\,f\!=\!0,1...N\,$  fermions and b\,=\,N-f bosons is
\begin{equation}G^0\,=\,G^0_1\,...\,G^0_f\,G^0_{f+1}\,...\,G^0_N,
\label{3}\end{equation} the propagators of the fermion i
and the boson j being respectively
\begin{equation}G^0_i\,=\,{1\over p_{i0}-h_i+i\epsilon
h_i}\,\beta_i\,,\label{4}\end{equation}
\begin{equation}G^0_j\,=\,{1\over
p^2_{j0}-E^2_j+i\epsilon}\,=\,{1\over2E_j}\,\sum_{\sigma_j}\, {\sigma_j\over
p_{j0}-\sigma_jE_j+i\epsilon\sigma_j}\,,\label{5}\end{equation} where $\,p_i\,$
is the 4-momentum of particle i\,, and
\begin{equation}h_i = \vec \alpha_i\, . \vec p_i + \beta_i\, m_i\,,\qquad
E_i=(\vec p_i^{\,2}+m_i^2)^{1\over 2}\,,\qquad \sigma_j\!=\!\pm1. 
\label{6}\end{equation}   
We do not specify the reference
frame in which we write noncovariant quantities like $\,h_i\,$ or $\,E_i.\,$ Our
3D reduction will in fact be frame-dependent. Practically, we shall choose the
global rest frame. The kernel $\,K\,$ will be chosen in such a way that
equation (\ref{1}) gives by iterations the usual expansion of $\,G\,$ in
terms of Feynman graphs. It is related to the kernel $\,K'\,$ defined in the
same way with the dressed propagators
\begin{equation}G\,=\,G'^0\,+\,G'^0K'G\label{1*}\end{equation}
where $\,G'^0\,$ is the product of the dressed propagators
\begin{equation}G'^0_i\,=\,{1\over \gamma_i\cdot
p_i\,-\,m_i\,-\,\Sigma_i\,+i\epsilon}\qquad\hbox{for a
fermion,}\label{b*}\end{equation}
\begin{equation}G'^0_i\,=\,{1\over
p^2_{i0}-E^2_i-\Sigma_i+i\epsilon}\qquad\hbox{for a
boson,}\label{c*}\end{equation}
and $\,\Sigma_i\,$ the renormalized self-energy function. The transfer of the
self-energies from the propagator to the kernel is achieved with \cite{20,21}   
\begin{equation}K\,=\,K'\,+\,\Sigma,\qquad\Sigma\,=\,(G^0)^{-1}-(G'^0)^{-1}.
\label{dd}\end{equation} 
In the two-body problem $\,K'\,$ is simply the sum of all irreducible Feynman
graphs. In the three-body problem, it is
\begin{equation}K'\,=\,K'_{12}\,(G'^0_3)^{-1}\,+\,
K'_{23}\,(G'^0_1)^{-1}\,+\,K'_{31}\,(G'^0_2)^{-1}\,+\,
K'_{123}\,,\label{7}\end{equation}
where $\,K'_{ij}\,$ is the sum of the two-body (ij) irreducible Feynman
graphs while $\,K'_{123}\,$ is the sum of the three-body connected
irreducible Feynman graphs. In the four-body problem and beyond, there
appear mutually unconnected kernels in the expression of $\,K'\,$ (like
$\,K'_{12}\,$ and $\,K'_{34}\,$ for example), and we shall have to be
careful to avoid overcountings in the expansion of $\,G\,$ (see next
section).\par
In ref. \cite{20}, we built a 3D reduction of the two-fermion Bethe-Salpeter
equation around an instantaneous approximation of the Bethe-Salpeter kernel, the
difference with the exact kernel being corrected in a series of higher-order
contributions to the final 3D potential. This potential being not explicitly
hermitian, we performed a second symmetrizing series expansion at the 3D level.
The result of the combination of both series turned out to be independent of the
starting instantaneous approximation, and was also
obtainable directly by starting with an approximation of the free propagator,
based on integrals in the relative energies instead of the more usual
$\,\delta-$constraints. Furthermore, the method was easily generalizable to a
system of N particles, consisting of any combination of bosons and fermions. In
the present work, we shall however present our 3D reduction in yet another way.
Let us first define the 3D free propagator
$$ \int
dp_0\,G^0(p_0)\,\equiv\int\delta\,(P_0-\sum_{i=1}^Np_{i0})\,dp_{10}...
dp_{N0}\,G^0(p_{10},...p_{N0})$$ 
\begin{equation}=\,\,{(-2i\pi)^{N-1}\over\omega}\,\,\tau\,g^0\,\beta
\label{8}\end{equation} with
\begin{equation}g^0\,=\,{1\over P_0-S+i\epsilon P_0}\,,\qquad
S=E\,(\Lambda^+-\Lambda^-)\,,\qquad
\beta=\beta_1...\beta_f\,,\label{9}\end{equation}
\begin{equation}\Lambda^{\pm}\,=\Lambda_1^{\pm}\,...\,\Lambda_f^{\pm}
\,,\qquad
\Lambda_i^{\pm}={E_i\pm
h_i\over2E_i}\,,\qquad\tau=\Lambda^++(-)^{f+1}\Lambda^-\,,\label{10}\end{equation}
\begin{equation}E\,=\,\sum_{i=1}^N\,E_i\,,\qquad\omega\,=\,2^b\,E_{f+1}
\,...\,E_{N}\label{11}\end{equation} for $\,f\neq0,\,$ so that
\begin{equation}\tau\,g^0\,=\,{\Lambda^+\over
P_0-E+i\epsilon}\,+\,(-)^{f+1}\,{\Lambda^-\over
P_0+E-i\epsilon}.\label{12}\end{equation} When $\,f\!=\!0,\,$ (bosons only), we
have no $\,\tau\,$ and no $\,\beta\,$ (we can replace them by 1 in (\ref{8}))
and
\begin{equation}g^0\,=\,{1\over P_0-E+i\epsilon}\,-\,{1\over
P_0+E-i\epsilon}\,=\,{2E\over
P_0^2-E^2+i\epsilon}\,.\label{13}\end{equation}                 
Let us now define a 3D full propagator in the $\,\tau^2\,$ subspace, with
$\,\tau g^0\,$ as free limit: 
\begin{equation}g\,=\,{1\over(-2i\pi)^{N-1}}\,{\tau
\sqrt{\omega}}\int
dp'_0dp_0\,G(p'_0,p_0)\,\beta\,{\tau\sqrt{\omega}}.\label{17}\end{equation}
The integrations with respect to the relative energies preserve the positions
of the bound state poles of $\,G\,$ (it can also be shown that the physical
particle-particle, particle-bound state or bound state-bound state scattering
amplitudes are also preserved \cite{KS,20}).    If we write
$\,G\,$ as
\begin{equation}G\,=\,G^0+G^0\,T\,G^0\,,\qquad
T\,=\,K(\,1-G^0K\,)^{-1}\label{18}\end{equation}
we get
\begin{equation}g\,=\,\tau g^0+\tau g^0<T>\tau g^0\label{19}\end{equation}
with the definition
\begin{equation}<A>={1\over(-2i\pi)^{N-1}}\,{\tau^2
\sqrt{\omega}\over g^0}\int
dp'_0dp_0\,G^0(p'_0)A(p'_0,p_0)G^0(p_0)\,\beta\,{\tau^2
\sqrt{\omega}\over g^0}.
\label{20}\end{equation}
To the 3D full propagator
$\,g\,$ we can associate the 3D equation
\begin{equation}\phi\,=\,g^0\,\tau\,V\,\phi\label{21}\end{equation}
where $\,V\,$ is such that
\begin{equation}g^{-1}\,=\,\tau \,(g^0)^{-1}\,-\,V.\label{21b}\end{equation}
The comparison with (\ref{19}) gives   
\begin{equation}<T>\,=\,V(1-\tau g^0 V\,)^{-1}\label{22}\end{equation}
or, conversely
$$V=<T>(1+\tau g^0\!<T>)^{-1}=<K(1\!-\!G^0K)^{-1}>(\,1+\tau
g^0\!<K(1\!-\!G^0K)^{-1}>)^{-1}$$
$$=\,\,<K(1\!-\!G^0K)^{-1}(\,1+>\tau
g^0\!<K(1\!-\!G^0K)^{-1})^{-1}>$$
\begin{equation}=\,\,<K(\,1\!-\!G^0K+>\tau
g^0\!<K)^{-1}>\,=\,\,<K(\,1\!-\!G^RK)^{-1}>\,=\,<K^T>\label{22.1}\end{equation}
with 
\begin{equation}
K^T\,=\,K(1-G^RK)^{-1}\,=\,K+KG^RK+\cdots\label{23}\end{equation}
\begin{equation}G^R\,=\,G^0-\,G^I\,,\qquad
G^I=\,\,\,>\tau g^0\!<\label{24}\end{equation} 
or, making the dependence on the relative energies explicit:
\begin{equation}
G^R(p'_0,p_0)=G^0(p_0)\,\delta\,(p'_0-p_0)-G^I(p'_0,p_0)\label{25}\end{equation}
with\begin{equation}
G^I(p'_0,p_0)\,=\,G^0(p'_0)\,\beta\,{\tau\,\omega
\over(-2i\pi)^{N-1}\,g^0}\,G^0(p_0).\label{25b}\end{equation} 
By substracting $\,G^I\,$ from $\,G^0,\,$ we
remove the leading terms which come from the residues of the positive-energy
poles of
$\,G^0.\,$ This ensures in principle the decreasing of the terms of the
series (\ref{23}).\par
The choice (\ref{17})  of $\,g\,$ as the integral of $\,G\,$ between two
$\,\tau\,$ operators defines our 3D reduction (in the two-body case it is also
Phillips and Wallace's reduction). A different approach, in the two-body case,
the constraint approach, consists in defining
$\,<T>\,$ by fixing the relative-energy arguments of $\,T(p'_0,p_0),\,$ for
example by putting one initial particle and one final particle 
on their positive-energy mass shell \cite{16}. There is no problem with this
approach for
$\,N\!=\!2.\,$ For $\,N\!=\!3,\,$ we should put two initial and two final
particles on their positive-energy mass shell. For the two-body
terms, this would give a constraint too much. This difficulty can be avoided by
applying different constraints to the different terms of
$\,T\,$ \cite{23,24}. This leads to a set of coupled equations which can
not, without approximations, be reduced to a single equation.      
\par   In \cite{20}, we derived equation (\ref{21}) from the inhomogeneous
Bethe-Salpeter equation (\ref{2}), with the definition  
\begin{equation}\phi\,=\,\tau\,\sqrt{\omega}\int
dp_0\,\Phi(p_0)\label{26}\end{equation}
 A simpler reduction method, which we
shall adopt from now, can be obtained by keeping only the positive-energy part
$\,\Lambda^+\,$ of
$\,\tau.\,$ In this case, the equation obtained after the reduction of the
Dirac spinors into Pauli spinors will be \cite{20} 
\begin{equation}(P_0-E\,)\,\varphi\,=\,\left[\,\prod_{i=1}^f\sqrt{2\,E_i\over
E_i+m}\,\,\right]\,\,v\,
\,\left[\,\prod_{i=1}^f\sqrt{2\,E_i\over
E_i+m}\,\,\right]\,\,\varphi.\label{27}\end{equation}
where $\,v\,$ is the large-large part of $\,V.$ This simpler
reduction method can be defined by the choice of $\,g\,$ as
the integral of $\,G,\,$ now between two positive-energy
projectors $\,\Lambda^+\,$ (instead of $\,\tau\,$ in
(\ref{17})). This 3D full propagator is, in configuration
space, the retarded part of $\,G\,$ taken at equal times.
It has been the starting point of a 3D reduction by
Logunov and Tavkhelidze \cite{4}  (in the two-fermion
case), followed by Kvinikhidze and Stoyanov \cite{KS} 
(three fermions) and Khvedelidze and Kvinikhidze \cite{26} 
(four fermions).                       

\section{The four-body problem.}
In this section 3, we shall neglect, for simplicity, the radiative corrections
(omitting thus the $\,\Sigma's\,$ and all the primes in the expression of
$\,K).$  
\subsection{Bethe-Salpeter equation for four particles.}
For four particles, we have \cite{25,26,27} 
\begin{equation}G^0\,=\,G^0_1\,G^0_2\,G^0_3\,G^0_4\label{28}\end{equation}
$$K\,=\,K_{12,34}\,+\,K_{13,24}\,+\,K_{14,23}$$
$$+\,\,\,K_{123}\,(G^0_4\,)^{-1}\,+\,K_{124}\,(G^0_3\,)^{-1}\,+\,K_{134}\,
(G^0_2\,)^{-1}\,+\,K_{234}\,(G^0_1\,)^{-1}\,$$ 
\begin{equation}+\,\,\,K_{1234}\,,\label{29}\end{equation} with
\begin{equation}K_{12,34}\,=\,K_{12}\,(G^0_3\,G^0_4\,)^{-1}
\,+\,K_{34}\,(G^0_1\,G^0_2\,)^{-1}\,-\,K_{12}\,K_{34}\label{30}\end{equation}
and similarly for $\,K_{13,24}\,$ and $\,K_{14,23}.\,$ The three lines of
(\ref{29}) contain the two, three and four-body kernels respectively. The last
term of (\ref{30}) is a counterterm which has no equivalent in the two
and three-body problems. If we write the expansion of $\,G$
\begin{equation}G\,=\,G^0\,+\,G^0\,K\,G^0\,+\,G^0\,K\,G^0\,K\,G^0
\,+\cdots\label{31}\end{equation}
and collect the terms containing one $\,K_{12}\,$ with one $\,K_{34},\,$ we
get indeed
$$G\,=\,\cdots\,-\,G^0K_{12}K_{34}G^0
\,+\,\,G^0K_{12}(G^0_3\,G^0_4\,)^{-1}G^0K_{34}(G^0_1\,G^0_2\,)^{-1}G^0$$
\begin{equation}
+\,\,G^0K_{34}(G^0_1\,G^0_2\,)^{-1}G^0K_{12}(G^0_3\,G^0_4\,)^{-1}G^0
\,+\,\cdots\label{32}\end{equation}
Since $\,K_{12}\,$ and $\,K_{34}\,$ commute mutually, the two last terms
of (\ref{32}) correspond to the same graph which appears thus twice,
while the first term is again the same graph with a minus sign. This
mechanism works at all orders. The kernel $\,K_{12,34}\,$ is indeed
given by
\begin{equation}K_{12,34}\,=\,(G^0\,)^{-1}-(G_{12,34})^{-1}\label{33}
\end{equation}  
where $\,G_{12,34}\,$ is the full propagator obtained by keeping only
the interactions $\,K_{12}\,$ and $\,K_{34}.\,$ It factorizes into
\begin{equation}(G_{12,34})^{-1}\,=\,(G_{12})^{-1}(G_{34})^{-1}
\,=\,\left[\,(G^0_1\,G^0_2\,)^{-1}-K_{12}\,\right]
\,\left[\,(G^0_3\,G^0_4\,)^{-1}-K_{34}\,\right]\label{34}\end{equation}
so that (\ref{33}) gives (\ref{30}). This problem of commutating
kernels was not present in the two and three-body problems. It has
been easily overcomed here in the four-body problem. It will
become much more complicated in the five-body problem and beyond \cite{25,27}.

\subsection{3D reduction of the four-body Bethe-Salpeter equation.}
For definiteness we shall work on  the Bethe-Salpeter equation for
four fermions and perform the reduction based on the positive-energy part of
$\,g^0.\,$ Then:
\begin{equation}\Lambda^+\,g^0\,=\,{\Lambda^+\over
P_0-E+i\epsilon}\label{35}\end{equation}
\begin{equation}<A>\,=\,{1\over(-2i\pi)^3}\,\Lambda^+(P_0-E\,)\int
dp'_0dp_0\,G^0(p'_0)A(p'_0,p_0)G^0(p_0)\,\beta\,\Lambda^+(P_0-E\,)
\label{36}\end{equation}
and the 3D equation will be
\begin{equation}\phi\,=\,g^0\,V\,\phi\label{37}\end{equation}
with
$$V\,=\,<K^T>\,=\,<K>+<KG^RK>+\cdots$$
$$=\,<K>+<K(G^0->g^0\!<)K>+\cdots$$
\begin{equation}
=\,\,<K>+<KG^0K>-<K>g^0<K>+\cdots\label{38}\end{equation}                 
We shall now compute the two and four-vertex terms, keeping only the
two-fermion kernels:
$$<K^T>^{(4)}\,=\qquad<K_{12,34}>\,+\,\cdots \quad(3\,\,terms)$$
$$+\left[\,<K_{12,34}\,G^0K_{14,23}>\,-\,\,<K_{12,34}>\,g^0<K_{14,23}>\,\right]\,+\,\cdots
\quad(6\,\,terms)$$
\begin{equation}+\left[\,<K_{12,34}\,G^0K_{12,34}>\,-\,\,<K_{12,34}>
\,g^0<K_{12,34}>\,\right]+\cdots
\quad(3\,\,terms)\label{39}\end{equation}
where
\begin{equation}<K_{12,34}>=<K_{12}\,(G^0_3\,G^0_4\,)^{-1}>
+<K_{34}\,(G^0_1\,G^0_2\,)^{-1}>-<K_{12}\,K_{34}>\label{40}\end{equation}
$$<K_{12,34}\,G^0K_{14,23}>^{(4)}\,=\,
<K_{12}\,(G^0_3\,G^0_4\,)^{-1}G^0K_{23}\,(G^0_1\,G^0_4\,)^{-1}>$$
$$
+<K_{12}\,(G^0_3\,G^0_4\,)^{-1}G^0K_{14}\,(G^0_2\,G^0_3\,)^{-1}>
+\,
<K_{34}\,(G^0_1\,G^0_2\,)^{-1}G^0K_{23}\,(G^0_1\,G^0_4\,)^{-1}>$$
\begin{equation}
\,+\,<K_{34}\,(G^0_1\,G^0_2\,)^{-1}G^0K_{14}\,(G^0_2\,G^0_3\,)^{-1}>
\label{41}\end{equation}
$$<K_{12,34}\,G^0K_{12,34}>^{(4)}\,=\,
<K_{12}\,(G^0_3\,G^0_4\,)^{-1}G^0K_{12}\,(G^0_3\,G^0_4\,)^{-1}>$$
$$
\,+\,<K_{34}\,(G^0_1\,G^0_2\,)^{-1}G^0K_{34}\,(G^0_1\,G^0_2\,)^{-1}>
+\,
<K_{12}\,(G^0_3\,G^0_4\,)^{-1}G^0K_{34}\,(G^0_1\,G^0_2\,)^{-1}>$$
\begin{equation}
\,+\,<K_{34}\,(G^0_1\,G^0_2\,)^{-1}G^0K_{12}\,(G^0_3\,G^0_4\,)^{-1}>
\label{42}\end{equation}  
The various terms of (\ref{40})-(\ref{42}) are represented in figure 1. The
two last terms of (\ref{42}) are equal to $\,<K_{12}\,K_{34}>.\,$ In the
expansion of $\,G\,$ this term will thus appear only once, with a plus
sign.\par
Let us now examine the counterterms $\,-<K>\!g^0\!<K>,\,$ which could be
represented by the 8 last graphs of figure 1, with a vertical line
separating them into two parts. In the two and three-particle problems we
have a cancellation between the corresponding contributions of
$\,<KG^0K>\,$ and  of \break$\,-<K>\!g^0\!<K>,\,$ when the two-body kernels
are approximated by (relative and total) energy independent kernels
and the propagators by positive-energy
propagators. This cancellation of the leading terms leads in principle to a
decreasing of the contributions of the higher-order terms of the series
$\,<K>\!+\!<KG^RK>\!+\cdots.\,$ We shall check this cancellation here in
(\ref{39}) by assuming that the two-body kernels contain positive-energy
projectors and are independent on the energies ($\,K_{ij}\!=\!\Lambda^+_i
\Lambda^+_jK_{ij}\beta_i\beta_j\Lambda^+_i\Lambda^+_j\,$ and is independent
of $\,p_{0ij},P_{0ij}\,$ -- we shall speak of the "positive-energy
instantaneous approximation", although "instantaneous" usually refers only
to the independence on the two-body relative energy). 
It is then easy to verify that the
cancellation occurs for the 4th to the 9th graphs of figure 1.
 We remain thus
with
$$<K^T>^{(4)}\,=\quad <K_{12}\,(G^0_3\,G^0_4\,)^{-1}>
\,+\,<K_{34}\,(G^0_1\,G^0_2\,)^{-1}>\,+\cdots$$
$$+\,<K_{12}\,K_{34}>\,-\,
<K_{12}\,(G^0_3\,G^0_4\,)^{-1}>g^0<K_{34}\,(G^0_1\,G^0_2\,)^{-1}>$$
\begin{equation}
\,-\,<K_{34}\,(G^0_1\,G^0_2\,)^{-1}>g^0<K_{12}\,(G^0_3\,G^0_4\,)^{-1}>
\,+\,\cdots\label{43}\end{equation} 
The two-vertex contributions become simply the sum of the six two-fermion
potentials:
\begin{equation}V\,=\,V_{12}\,+\,V_{34}\,+\,V_{13}\,+
\,V_{24}\,+\,V_{14}\,+\,V_{23}\,,\label{44}\end{equation}
\begin{equation}V_{ij}\,=\,-2i\pi\,\beta_i\beta_j\,K_{ij}.
\label{45}\end{equation}
For $\,<K_{12}\,K_{34}>,\,$ we have
$$<K_{12}\,K_{34}>\,=\,{-1\over2i\pi}\int dP_{120}\,dP_{340}\,\,
\delta\,(P_0-P_{120}-P_{340}) $$
$$(P_0-E')\,\left[\,{1\over P_{120}-E'_1-E'_2+i\epsilon}\,\,V_{12}\,\,
{1\over P_{120}-E_1-E_2+i\epsilon}\,\right]$$
\begin{equation}\left[\,{1\over P_{340}-E'_3-E'_4+i\epsilon}\,\,V_{34}\,\,
{1\over P_{340}-E_3-E_4+i\epsilon}\,\right]\,(P_0-E).\label{46}\end{equation}
Let us perform the integration with respect to $\,P_{120}\,$ and close the
integration path clockwise. We have to consider the poles at
$\,P_{120}\!=\!E_1\!+\!E_2\,$ and $\,P_{120}\!=\!E'_1\!+\!E'_2.\,$ We obtain
$$<K_{12}\,K_{34}>\,=\,{P_0-E'\over
E_1+E_2-E'_1-E'_2}\,\,V_{12}\,\, {1\over
P_{0}-E_1-E_2-E'_3-E'_4}\,\,V_{34}$$
\begin{equation}+\,\,V_{12}\,\,{1\over
E'_1+E'_2-E_1-E_2}\,\,V_{34}\,\, {P_0-E\over
P_{0}-E'_1-E'_2-E_3-E_4}\,.\label{47h}\end{equation}
Writing then
\begin{equation}P_0-E'\,=\,(P_0-E_1-E_2-E'_3-E'_4\,)\,+\,
(E_1+E_2-E'_1-E'_2\,)\label{48}\end{equation}
\begin{equation}P_0-E\,=\,(P_0-E'_1-E'_2-E_3-E_4\,)\,+\,
(E'_1+E'_2-E_1-E_2\,)\label{49}\end{equation} 
in (\ref{47h}), we see that the contributions of the first terms cancel
mutually, so that we remain with
\begin{equation}<K_{12}K_{34}>=V_{12}V_{34}\left[ {1\over
P_{0}-E_1-E_2-E'_3-E'_4}+{1\over
P_{0}-E'_1-E'_2-E_3-E_4}\right].\label{47}\end{equation} 
These two terms are cancelled by the two last terms of (\ref{43})
respectively.\par
We have thus seen that, in the case of instantaneous positive-energy kernels,
the contribution of the four-vertex terms vanishes, while the contribution of the
two-vertex terms is simply the sum of the six two-fermion potentials. Let us
 now examine a typical set of six-vertex graphs, the ones which contain two
$\,K_{12}\,$ with one
$\,K_{34}.\,$ Their contributions to the potential are represented in figure 2,
with a vertical line when $\,G^0\,$ is to be replaced by $\,>g^0\!<.\,$ 
After
the mutual cancellations of some identical contributions, we obtain the
corresponding contribution to
$\,<T>,\,$ minus the four graphs containing one $\,>g^0\!<,\,$
plus the three graphs containing two  $\,>g^0\!<.\,$ Let us
now consider again the case of instantaneous potentials with
positive-energy propagators. We have shown above that, in a sequence
$\,K_{12}\,K_{34},\,$ we must consider the two possible orders. Here, in a
sequence $\,K_{12}\,K_{12}\,K_{34},\,$ it can be shown that we must
consider the three possible orders, so that the first graph of figure 2, e.g.,
becomes equal to the sum of the three last ones. In figure 3, we expand
the graphs of figure 2 in the case of instantaneous kernels with
positive-energy propagators, and we see that the sum of the resulting
graphs is zero.

\section{Bypassing the Bethe-Salpeter equation.}
In section 3, we presented a 3D reduction method for the N-body Bethe-Salpeter
equation. The starting homogeneous equation was written in terms of a kernel
$\,K,\,$ which was designed to reproduce the full propagator $\,G\,$ by
iterations of the inhomogeneous equation. We have seen that the writing of this
kernel $\,K\,$ was straightforward for $\,N=2,3,\,$ less straightforward
for $\,N=4,\,$ and increasingly complicated for $\,N\ge5.\,$ Our
investigations on the $\,N=4\,$ case in section 3 suggests us a possible
way of avoiding the explicit writing of the Bethe-Salpeter equation. We saw in
section 2 that the 3D potential is given by
\begin{equation}V\,=\,<T(1+G^IT\,)^{-1}>\,=\,<K(1-G^RK\,)^{-1}>\label{b1}
\end{equation}
where $\,G^R\!=\!G^0\!-\!G^I,\,$ while $\,K\,$ contain counter-terms in order
to avoid the apparition of topologically identical diagrams in the
expansion of $\,T\!=\!K(1\!-\!G^0K)^{-1}.\,$ It is however possible to adopt
a simpler algorithm for $\,T:$
\begin{equation}T\,=\,\left[K^{IR}(1-G^0K^{IR})^{-1}
\right]^{SC}.\label{b2}\end{equation}
By $\,K^{IR}\,$ we denote the sum of the irreducible interactions (as
(\ref{29}-\ref{30}) without the three counter-terms) and by $\,SC\,$  (single
counting), we mean that the diagrams which appear two or more times in
the expansion of $\,T\,$ must be kept only once. The 3D potential will
then be written
\begin{equation}V\,=\,<K^{IR}(1-G^RK^{IR})^{-1}>^{SC}\label{b3}\end{equation}
in which, after the expansion of the series and the splitting of $\,G^R\,$ into
$\,G^0\!-\!G^I,\,$ we shall remove the duplicating diagrams. This
makes the writing of the 3D potential as straightforward for
$\,N\!\ge\!4\,$ as it was for $\,N\!=\,$1 or 2.  As an example in
the four-particle case:
$$(K_{12}G^RK_{34}\,+\,K_{34}G^RK_{12})^{SC}$$
$$=\,(K_{12}G^0K_{34}\,+\,K_{34}G^0K_{12}\,-
\,K_{12}G^IK_{34}\,-\,K_{34}G^IK_{12})^{SC}$$
\begin{equation}=\,K_{12}G^0K_{34}\,-
\,K_{12}G^IK_{34}\,-\,K_{34}G^IK_{12}.\label{b4}\end{equation}
With positive-energy instantaneous interactions, the term in
$\,G^0\,$ will be cancelled by the two terms in $\,G^I.$\par
The writing of the Bethe-Salpeter kernel could be done in a quite similar way:
\begin{equation}K\,=\,\left\{T
(1+\,G^IT)^{-1}\right\}_{G^I\to
G^0}\,=\,\left\{\left[K^{IR}
(1-[\,G^0\!-\!G^I]\,K^{IR})^{-1}\right]^{SC}\right\}_{G^I\to
G^0}.\label{b5}\end{equation}
Here, $\,G^I\,$ is simply a temporary renaming of $\,G^0,\,$ which indicates
that terms like $\,AG^IB\,$ and $\,BG^IA\,$ must always be kept both. We
know that
$\,K,\,$ unlike
$\,V,\,$ contain only a finite number of parts \cite{27}. For $\,N\!=\!4,\,$
e.g., eqs. (\ref{29}-\ref{30}) show that
$\,K\,$  contains 11 subkernels (groups of irreducible graphs) in $\,K^{IR}\,$
plus 3 counterterms, coming from the second order in $\,K^{IR}\,$ (we get one of
these counterterms by replacing $\,G^I\,$ by $\,G^0\,$ in (\ref{b4})). 
\par The operator $\,T\,$ is directly given by the Feynman graphs if
we neglect the radiative corrections. If not, we can use
\begin{equation}G^0+G^0TG^0\,=\,G\,=\,G'^0+G'^0T'G'^0\label{aa1}\end{equation}
with
\begin{equation}G'^0\,=\,G^0\,(1-\Sigma\,G^0\,)^{-1}\,=\,
\prod_i G^0_i\,(1-\Sigma_i\,G^0_i\,)^{-1}.
\label{aa2}\end{equation}
The operators $\,G\,$ and $\,T'\,$ are directly given by the Feynman graphs.
For $\,T,\,$ we have   
\begin{equation}T\,=\,\Sigma(1\!-\!G^0\Sigma)^{-1}\,+\,(1\!-\!\Sigma
G^0)^{-1}\,T'
\,(1\!-\!G^0\Sigma)^{-1}.
\label{aa3}\end{equation}

\section{Conclusions}
In a previous work \cite{20}, we were in search of a 3D reduction method for
the two-fermion Bethe-Salpeter equation based on an unspecified positive-energy
instantaneous approximation of the Bethe-Salpeter kernel. We performed a series
expansion around this approximation, followed by an integration on the relative
energy and a second series expansion, at the 3D level, in order to render the
resulting 3D potential symmetric. After combining both series, we found that we
had in fact built a kind of propagator-based reduction, using an integration
with respect to the relative energy, in full contrast with the usual
constraining propagator-based reductions, which use constraints. Furthermore,
this method was easily generalisable to systems consisting in any number of
fermions and/or bosons.\par
In the present work, the increasing difficulty of writing the Bethe-Salpeter
kernel for $\,N\ge4\,$ suggested us a direct way of writing the 3D potential
without the need of first writing the Bethe-Salpeter kernel explicitly. This
writing of the 3D potential is straightforward and valid for all $\,N,\,$ with
the simple prescription of removing the duplicating graphs which appear when
$\,N\ge4.,$

\end{document}